# Science Models as Value-Added Services for Scholarly Information Systems

Peter Mutschke, Philipp Mayr, Philipp Schaer, York Sure

*GESIS - Leibniz-Institute for the Social Sciences,
Lennéstr. 30, 53111 Bonn, Germany*

Phone:   +49 228 2281 135

Fax:      +49 228 2281 120

E-mail:  peter.mutschke@gesis.org

The paper introduces scholarly Information Retrieval (IR) as a further dimension that should be considered in the science modeling debate. The IR use case is seen as a validation model of the adequacy of science models in representing and predicting structure and dynamics in science. Particular conceptualizations of scholarly activity and structures in science are used as value-added search services to improve retrieval quality: a co-word model depicting the cognitive structure of a field (used for query expansion), the Bradford law of information concentration, and a model of co-authorship networks (both used for re-ranking search results). An evaluation of the retrieval quality when science model driven services are used turned out that the models proposed actually provide beneficial effects to retrieval quality. From an IR perspective, the models studied are therefore verified as expressive conceptualizations of central phenomena in science. Thus, it could be shown that the IR perspective can significantly contribute to a better understanding of scholarly structures and activities.

## Introduction

Science models usually address issues in statistical modeling and mapping of structures and scholarly activities in science. As a further dimension, that should be considered in science modeling as well, the paper focuses on the applicability of science models in scholarly Information Retrieval (IR) with regard to the improvement of search strategies in growing scientific information spaces. Introducing an IR perspective in science modeling is motivated by the fact that scholarly IR as a science of searching for scientific content can be also seen as a special scholarly activity that therefore should also be taken into account in science modeling. Moreover, as scholarly Digital Libraries (DLs) can be considered as particular representations of the science system, searching in DLs can be seen as a particular use case of interacting with exactly that system that is addressed by science modeling. From this perspective, IR can play the role of a validation model of the science models under study.

From the perspective of IR, a further motivation point is the assumption that traditional IR approaches fail at points where the application of science models may help: (1) the vagueness between search and indexing terms, (2) the information overload by the amount of result records obtained, and (3) the problem that pure term frequency based rankings provide results that often do not meet user needs (Mayr et al. 2008). This strongly suggests the introduction of science models in IR systems that rely more on the real research process and have therefore a greater potential for closing the gap between information needs of scholarly users and IR systems than conventional system-oriented approaches. While, in this paper we mainly focus on how to use science model-enhanced IR as a test bed for different science models, we would also like to point out that there is a further interface between IR and scientometrics which is currently underexploited. One of the problem solving tasks shared by IR and scientometrics is the determination of a "proper" selected set of documents from an ensemble. In particular for newly emerging interdisciplinary fields and their evaluation the definition of the appropriate reference set of documents is important. Glänzel et al. (2009) have discussed how bibliometrics can be also used for the retrieval of "core literature". Bassecoulard et al. (2007) and Boyack & Klavans (2010) proposed sophisticated methods to delineate fields on the basis of articles as well as journals. However, due to the interconnectedness of research streams and different channels of knowledge transfer, it remains a complex problem how "hard boundaries" in continuously changing research landscapes can be found.

In their paper on Bibliometric Retrieval Glänzel et al. (2009) apply a combination of methods. They start from bibliographic coupling and keyword-based search and continue with a step-wise process to filter out the final core set from potentially relevant documents. Hereby, they make use of methods that are standard techniques in traditional IR as well (such as keyword-based search or thresholds). But, as already stated by Glänzel et al., "the objectives of subject delineation in the framework of bibliometric (domain) studies essentially differ from the goals of traditional information retrieval". In principle, this requires the application of different methods.

The bibliometric retrieval approach, in particular in an evaluative context, aims at defining a reference set of documents on the basis of a firm methodological canon, in order to justify the application and interpretation of standardized indicators. In traditional IR, in contrast, the application of bibliometric models and approaches has the primary goal to enhance the search from the perspective of the user by combining a wider search space with a particular contextualization of the search. The overall aim here is to help the user to get a grasp about the size and structure of the information space, rather than forcing him to precisely define the search space.

Correspondingly, the goal of the DFG-funded project "Value-added Services for Information Retrieval"[1] (Mayr et al. 2008) presented in this paper therefore is to improve retrieval quality in scholarly information systems by computational science models that reason about structural properties of the science system under study. Accordingly, the overall assumption of the IRM project is that a user's search should improve when science model driven search services are used. The other way around, the extent to which retrieval quality can be improved by performing science models as search services is seen as an indicator for the adequacy of the models taken in representing and predicting scholarly activities in the science system under study.

In the following, we will at first introduce the models proposed. After that, the evaluation study is presented. The paper closes with a discussion of the observed results and the conclusions to be drawn for the models studied.

## Models

Computational science models, to our understanding, are particular conceptualizations of scholarly activities and structures that can be expressed in algorithms (to be operationalized in systems that – more or less - reason about science, such as IR systems). The paper proposes three different kinds of science models as value-added search services that highlight different aspects of scholarly activity (see Figure 1): (1) a co-word model of science addressing the cognitive structure of a field by depicting the relationships between terms in a field (STR), (2) a bibliometric model of re-ranking, called Bradfordizing, representing the publication form of research output and its organization on a meso-level in terms of journals (BRAD), and (3) a co-authorship model of re-ranking examining the collaboration between the human actors of knowledge flow in science (AUTH). STR addresses the problem of the vagueness between search and indexing terms by pointing to co-related terms that are more appropriate for searching, BRAD and AUTH the problem of large and unstructured result sets by ranking a given document set according to the coreness of journals (BRAD) or according to the centrality of authors (AUTH) in a scientific community. Thus, the three models adress very different dimensions of structural properties in the science system. Moreover, they are also heterogeneous as regards the methods applied. The STR uses co-word analysis, BRAD bibliometric statistics, and AUTH methods taken from social network analysis, graph theory respectively.

---

[1] http://www.gesis.org/irm/

However, to the same extent as different science models emphasize different aspects of scholarly activity we expect that different kind of searches are best served by relying on corresponding science models. This approach meets the fact that the frequency of many structural attributes of the science system (such as co-authorships) usually follows some type of power law distribution. These highly frequent attributes which are produced when applying the science models have a strong selectivity in the document space which can be utilized for IR. In the following the three models, which are descriptive models of science so far, are discussed on a general conceptual level.

**A co-word model of science: relevance of co-terms**

Search in textual information systems only works when a user can find the right search terms describing his information need and the terms used in the information system. This mapping problem is known as the Language Problem in IR (Blair 1990, 2003). While formulating queries a user is in an "Anomalous State of Knowledge" (Belkin 1980) – trying to map the words and concepts describing his problem to the terms of the system fighting ambiguity and vagueness of language. This problem especially occurs in highly specialized scientific literature databases where often only literature reference with spare bibliographic metadata is available. Another source of vagueness evolves from special discourse dialects in scientific communities. These dialects are not necessarily the same dialects an information specialist would use to describe a document or a concept using his documentation language–.

Therefore, an instrument is needed to map the user's query terms to the document terms of the system. Especially in digital libraries searchers are confronted with databases that contain merely short texts which are described with controlled vocabularies. User studies in digital libraries have shown that most users are not aware of the special controlled vocabularies used in digital libraries (Shiri and Revie 2006). Hence they are not using them in their query formulation.

Co-word analysis (Callon et al. 1983) can be used to reduce the problem of language ambiguity and the vagueness of query. Petras (2006) proposed a search term suggestion system (STR) which relies on a co-word model of science that maps query terms to indexing terms at search time on the basis of term-term-associations between two vocabularies: (1) natural language terms from titles and abstracts and (2) controlled vocabulary terms used for document indexing. The associations are weighted according to their co-occurrence within the collection to predict which of the controlled vocabulary terms best mirror the search terms (Plaunt and Norgard 1998, Buckland 1999). The weights are calculated with the aid of a contingency table of all possible combinations of the two terms *A* and *B*: $AB$, $A\neg B$, $\neg AB$ and $\neg A\neg B$ where "$\neg$" denotes the absence of a term. Indexing terms having a high association with a query term are then recommended to the user for searching.

Thus, the science model proposed focuses on the cognitive structure of a field depicting the cognitive contexts in which a term appears. Accordingly, highly associated terms are not just related terms or synonyms. Terms that strongly appear together (in the sense of the model) rather represent the cognitive link structure of a field, i.e. they represent the co-issues that are discussed together within the research context in question. Thus, the STR is not a dictionary pointing to related terms. To what the STR really points are scientific discourses in which the user's term appears such that the user is provided by the research issues related to his/her term, i.e. the cognitive structure of the field in which the initial term is embedded. In the information system this cognitive structure is described by a controlled vocabulary, used systematically for indexing the documents in the system, such that a high probability of precise retrieval results is expected when these terms are used (instead of natural language terms of the user). In an IR environment a STR can be used as a query expansion mechanism by enriching the original query with highly relevant controlled terms derived from the special documentation language. Query expansion is the process of reformulation an initial query to improve retrieval performance in an information retrieval process (Efthimiadis 1996) and can be done in two ways: manually/interactively or automatically. Done interactively this kind of reformulation help may improve the search experience for the user in general. Suggesting terms reduces the searcher's need to think of the right search terms that might describe his or her information need. It effectively eases the cognitive load on the searcher since it is much easier for a person to pick appropriate search terms from a list than to come up with search terms by themselves (White and Marchionini 2007).

A further effect of the STR is that it may point the user to different expressions for the concept the user has in mind. A new or different view on a topic may ease the user to change the search strategy towards related issues of the field (which are represented in the cognitive structure the STR is providing). Thus, in an interactive scenario suggested terms or concepts can even help to alleviate "anchoring bias" (Blair 2002) which describes the human tendency to rely too heavily on one concept or piece of information when making decision. This cognitive effect can be worked against by suggesting terms and encourage a variation in one's initial search strategy and a reconsideration on the query formulation.

**A bibliometric model of science: coreness of journals**

Journals play an important role in the scientific communication process (cp. Leydesdorff et al. 2010). They appear periodically, they are topically focused, they have established standards of quality control and often they are involved in the academic gratification system. Metrics like the famous impact factor are aggregated on the journal level. In some disciplines journals are the main place for a scientific community to communicate and discuss new research results. These examples shall illustrate the impact journals bear in the context of science models.

Modeling science or understanding the functioning of science has a lot to do with journals and journal publication characteristics. These journal publication characteristics are the point where Bradford law can contribute to the larger topic of science models.

Bradford law of scattering bases on literature observations the librarian S. Bradford has been carried out in 1934. His findings and after that the formulation of the bibliometric model stand for the beginning of the modern documentation (Bradford 1948) – a documentation which founds decisions on quantifiable measures and empirical analyses. Bradford's work bases on analyses with journal publications on different subjects in the sciences.

Fundamentally, Bradford law states that literature on any scientific field or subject-specific topic scatters in a typical way. A core or nucleus with the highest concentration of papers - normally situated in a set of few so-called core journals - is followed by zones with loose concentrations of paper frequencies (see Figure 1 for a typical Bradford distribution). The last zone covers the so-called periphery journals which are located in the model far distant from the core subject and normally contribute just one or two topically relevant papers in a defined period. Bradford law as a general law in informetrics can successfully be applied to most scientific disciplines, and especially in multidisciplinary scenarios (Mayr 2009). Bradford describes his model in the following:

"The whole range of periodicals thus acts as a family of successive generations of diminishing kinship, each generation being greater in number than the preceding, and each constituent of a generation inversely according to its degree of remoteness." (Bradford 1934)

Bradford provides in his publications (1934, 1948) just a graphical and verbal explanation of his law. A mathematical formulation has been added later by early informetric researchers. Bradford`s original verbal formulation of his observation has been refined by Brookes (1977) to a cumulative distribution function resulting in a so-called rank-order distribution of the items in the samples. In the literature we can find different names for this type of distribution, e.g. "long tail distribution", "extremely skewed", "law of the vital few" or "power law" which all show the same properties of a self-similar distribution. In the past, Bradford law is often applied in bibliometric analyses of databases and collections e.g. as a tool for systematic collection management in library and information science. This has direct influence on later approaches in information science, namely the development of literature databases. The most common known resource which implements Bradford law is the Web of Science (WoS). WoS focuses very strictly on the core of international scientific journals and consequently neglects the majority of publications in successive zones.

Bradfordizing, originally described by White (1981), is a simple utilization of the Bradford law of scattering model which sorts/re-ranks a result set accordingly to the rank a journal gets in a Bradford distribution. The journals in a search result are ranked by the frequency of their listing in the result set, i.e. the number of articles in a certain journal. If a search result is "bradfordized", articles of core journals are ranked ahead of the journals which contain only an average number (Zone 2) or just few articles (Zone 3) on a topic (compare the example in Figure 1). This re-ranking method is interesting because it is a robust and quick way of sorting the central publication sources for any query to the top positions of a result set such that "the most productive, in terms of its yield of hits, is placed first; the second-most productive journal is second; and so on, down through the last rank of journals yielding only one hit apiece" (White 1981).[2]

Thus, Bradfordizing is a model of science that is of particular relevance also for scholarly information systems due to its structuring ability and the possibility to reduce a large document set into a core and succeeding zones. On the other hand, modeling science into a core (producing something like coreness) and a periphery always runs the risk and critic of disregarding important developments outside the core.

**A network model of science: centrality of authors**

The background of author centrality as a network model of science is the perception of "science (as) a social institution where the production of scientific knowledge is embedded in collaborative networks of scientists" (He 2009, see also Sonnewald 2007). Those networks are seen as "one representation of the collective, self-organized emerging structures in science" (Börner and Scharnhorst 2009). Moreover, because of the increasing complexity of nowadays research issues collaboration is becoming more and more "one of the key concepts in current scientific research communication" (Jiang 2008). The increasing significance of collaboration in science not only correlates with an increasing amount (Lu and Feng 2009, Leydesdorff and Wagner 2009) but also, and more importantly, with an increasing impact of collaborative papers (Beaver 2004, Glänzel et. al 2009 Lang and Neyer 2004).

---

[2] Bradfordizing can be applied to document types other than journal article, e.g. monographs (cf. Worthen 1975; Mayr 2008, 2009). Monographs e.g. provide ISBN numbers which are also good identifiers for the Bradfordizing analysis.

Collaboration in science is mainly represented by co-authorships between two or more authors who write a publication together. Transferred to a whole community, co-authorships form a co-authorship network reflecting the overall collaboration structure of a community. Co-authorship networks have been intensively studied. Most of the studies, however, focus mainly either on general network properties (see Newman 2001, Barabasi et al. 2002) or on empirical investigation of particular networks (Yin et al. 2006, Liu et al. 2005). To our knowledge, Mutschke was among the first who pointed to the relationship between co-authorship networks and other scientific phenomena, such as cognitive structures (Mutschke 1995, Mutschke and Quan-Haase 2001), and particular scientific activities, such as searching scholarly DLs (Mutschke 1994, 2001ff.).

From the perspective of science modeling it is important to note that, as co-authorships also indicate the share of knowledge among authors, "a co-authorship network is as much a network depicting academic society as it is a network depicting the structure of our knowledge" (Newman 2004). A crucial effect of being embedded in a network is that "some individuals are more influential and visible than others as a result of their position in the network" (Yin et al. 2006). As a consequence, the structure of a network also affects the knowledge flow in the community and becomes therefore an important issue for science modeling as well as for IR (cp. Mutschke and Quan-Haase 2001, Jiang 2008, Lu and Feng 2009, Liu et al. 2005).

This perception of collaboration in science corresponds directly with the idea of structural centrality (Bavelas 1948; Freeman 1977) which characterizes centrality as a property of the strategic position of nodes within the relational structure of a network. Interestingly, collaboration in science is often characterized in terms that match a particular concept of centrality widely used in social network analysis, namely the betweenness centrality measure which evaluates the degree to which a node is positioned *between* others on shortest paths in the graph, i.e. the degree to which a node plays such an intermediary role for other pairs of nodes. Yin et al. (2006) see co-authorship as a "process in which knowledge flows among scientists". Chen et al. (2009) characterize "scientific discoveries as a brokerage process (which) unifies knowledge diffusion as an integral part of a collective information foraging process". That brokerage role of collaboration correlates conceptually to the betweenness measure which also emphasizes the bridge or brokerage role of a node in a network (Freeman 1977, 1978/79, 1980, cp. Mutschke 2010).

The betweenness-related role of collaboration in science was confirmed by a number of empirical studies. Yan and Ding (2009) discovered a high correlation between citation counts and the betweenness of authors in co-authorship networks. Liu et al (2005) discovered a strong correlation between program committee membership and betweenness in co-authorship networks. Mutschke and Quan-Haase (2001) observed a high correlation of betweenness in co-authorship networks and betweenness of the author's topics in keyword networks. High betweenness authors are therefore characterized as "pivot points of knowledge flow in the network" (Yin et al. 2006). They can be seen as the main driving forces not only for just bridging gaps between different communities but also, by bringing different authors together, for community making processes.. This strongly suggests the use of an author centrality model of science also for re-ranking in scholarly IR (cf. Zhou et al. 2007). Our model of author centrality based ranking originates from the model initially proposed by Mutschke (1994) which has been re-implemented for a real-life IR environment, to our knowledge before anyone else, within the Daffodil system (Mutschke 2001, Fuhr et al. 2002) and the infoconnex portal (Mutschke 2004a,b). Currently, the ranking model is provided by the German Social Science portal sowiport[3] as a particular re-ranking service. The general assumption of the model is that a publication's impact can be quantified by the impact of their authors which is given by their centrality in co-authorship networks (cp. Yan and Ding 2009). Accordingly, an index of betweenness of authors in a co-authorship network is seen as an index of the relevance of the authors for the domain in question and is therefore used for re-ranking, i.e. , a retrieved set of publications is re-ranked according to the betweenness values of the publications' authors such that publications of central authors are ranked on top.

---

[3] www.gesis.org/sowiport

However, two particular problems emerge from that model. One is the conceptual problem of author name ambiguity (homonymy, synonymy) in bibliographic databases. In particular the potential homonymy of names may misrepresent the true social structure of a scientific community. The other problem is the computation effort needed for calculating betweenness in large networks that may bother, in case of long computation times, the retrieval process and finally user acceptance. In the following an evaluation of the retrieval quality of the three science mode driven search services are presented.

## Evaluation

### Proof-of-Concept Prototype

All three proposed models were implemented in an online information system[4] to demonstrate the general feasibility of the three approaches. The prototype uses those models as particular search stratagems (Bates 1990) to enhance retrieval quality. The open source search server Solr[5] is used as the basic retrieval engine which provides a standard term frequency based ranking mechanism (TF-IDF). All three models work as retrieval add-ons on-the-fly during the retrieval process. The STR module is based on a commercial classification software (Recommind Mindserver). The term associations are visualized as term clouds such that the user can see the contexts in which the terms the user has in mind appear in the collection. This enables the user to select more appropriate search terms from the cloud to expand the original query.

---

[4] www.gesis.org/beta/prototypen/irm
[5] http://lucene.apache.org/solr/

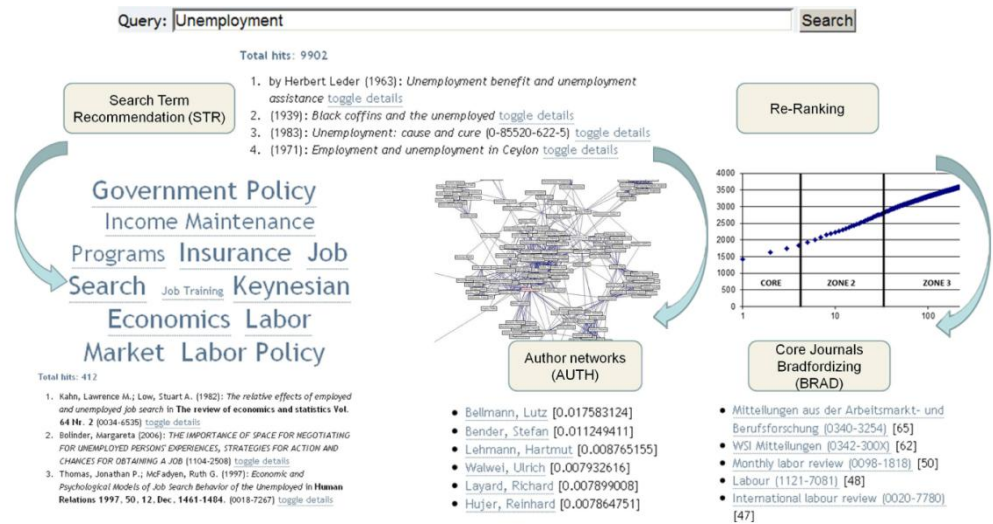

Figure 1: A simple search example (query term: "Unemployment") and typical structural attributes/outputs of implemented science models in our retrieval system. From left: Search Term Recommendation (STR) producing highly associated indexing terms, Author Networks (AUTH) with centrality-ranked author names and Bradfordizing based on Core Journals (BRAD) with highly frequent journal names/ISSNs.

The Bradfordizing re-ranking model is implemented as a Solr plugin which orders all search results with an ISSN number such that the journal with the highest ISSN count gets the top position in the result set, the second journal the next position, and so forth. The numerical TF-IDF ranking value of each journal paper in the result set is then multiplied with the frequency count of the respective journal. The result of this computation is taken for re-ranking such that core journal publications are ranked on the top.

The author centrality based re-ranking model computes a co-authorship network on the basis of the result set retrieved for a query, according to the co-authorships appearing in the result set documents[6]. For each node in the graph betweenness is measured, and each document in the result set is assigned a relevance value given by the maximum betweenness value of its authors. Single authored publications are captured by this method if their authors appear in the graph due to other publications they have published in co-authorship. Thus, just publications from pure single fighters are ignored by this procedure. The result set is then re-ranked by the centrality value of the documents' authors such that publications of central authors appear on the top of the list.

---

[6] Actually, the author-author-relations are computed during indexing time and are retrieved by the system via particular facets added to the user's query.

**Methods**

A major research issue of the project is the evaluation of the contribution of the three services studied to retrieval quality: Do central authors, core journals respectively, actually provide more relevant hits than conventional text-based rankings? Does a query expansion by highly associated co-words of the initial query terms have any positive effects on retrieval quality? Do combinations of the services enhance the effects? By measuring the contribution of our services to retrieval performance we expect deeper insights in the structure and the functioning the science system: As searching in a scientific information system is seen as a way of interacting with the science system, retrieval quality evaluation might also play the role of a "litmus test" for the adequacy of the science models taken for understanding, forecasting and communicating the science system. Thus, the investigation of science model driven value added services for scholarly information systems might contribute to a better understanding of science.

The standard approach to evaluate IR systems is to do relevance assessments i.e. the documents retrieved are marked as relevant or not relevant with respect to previously defined information need (cf. TREC[7], CLEF[8]). Modern collections usually are large and can't be assessed in total. Therefore, only subsets of the collection are assessed by a so-called pooling method (Voorhees and Harman 2005) where the top *n* documents returned by the different IR systems are taken. The assessors have to judge just the documents in the subsets without knowing the originating IR systems.

**Data and Topics**

In our study, a precision test of the three proposed models was conducted by performing user assessments of the top ten documents provided by each of the three services. As a baseline, the initial TF-IDF ranking done by the Solr engine was taken and therefore also judged. As regards STR, the initial query was expanded automatically by the four strongest associated co-words.

---

[7] http://trec.nist.gov/
[8] http://www.clef-campaign.org/

The precision test was carried out with 73 participants for ten different predefined topics from the CLEF corpus. Each assessor had to choose one out of the ten topics. The judgments were done according to the topic title and the topic description. The assessors were instructed how to assess in a briefing. Each of the four evaluated systems (AUTH = re-ranking by author centrality, BRAD = re-ranking by Bradfordizing, STR = TF-IDF ranking for the expanded query, and SOLR as the baseline) returned the $n$=10 top ranked documents, which formed the pool of documents to be assessed. Duplicates were removed, so that the size of the sample pools was between 34 and 39 documents. The assessors had to judge each document in the pool as relevant or not relevant (binary decision). In case they didn't assess a document this document is ignored in later calculations.

The retrieval test was performed on the basis of the SOLIS[9] database that consists of 369,397 single documents (including title, abstract, controlled keyword etc The 73 assessors did 43.78 single assessments on average which sums up to 3,196 single relevance judgments in total. Only 5 participants didn't fill out the assessment form completely, but 13 did more than one. Since every assessor could freely choose from these topics the assessments are not distributed evenly.

**Results**

*Overall agreement and Inter-grader Reliability*

The assessors in this experiment were not professionals and/or domain experts but students (mainly library and information science). However, according to findings in TREC where a rather high overall agreement between official TREC and non-TREC assessors was observed (Al-Maskari et al. 2008, Alonso and Mizzaro 2009), we also assume not a significant difference between domain experts and students in information science since all topics are every-day life topics[10]. The 73 assessors in our study judged a total of 3,196 single documents with an overall agreement over all topics and among all participants of 82%. 124 of 363 cases

---

[9] http://www.gesis.org/solis
[10] However, a retrieval study with experts from different domains is currently carried out.

where perfect matches where all assessors agreed 100% (all relevant and non relevant judgments matched). To rate the reliability and consistency of agreement between the different assessments we applied the Fleiss's Kappa measure of inter-grader reliability for nominal or binary ratings (Fleiss 1971). All Kappa scores in our experiment range between 0.20 and 0.52 which indicates a mainly acceptable level of overall agreement (see more details Schaer et al. 2010).

*Precision*

The precision *P* of each service was calculated by

$$P = \frac{|r|}{|r+nr|} \tag{12}$$

for each topic, where *|r|* is the number of all relevant assessed documents and *|r+nr|* is the number of all assessed documents (relevant and not relevant). A graph of all precision values including standard error can be seen in Figure 2.

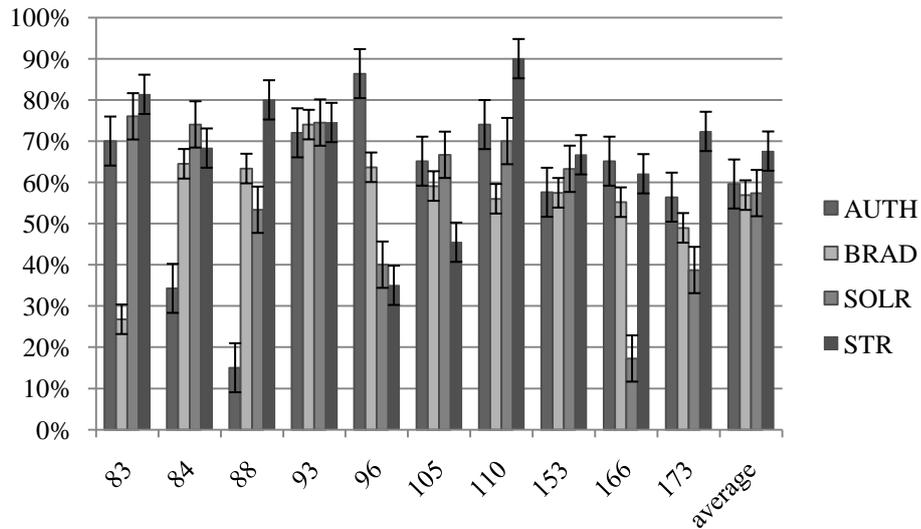

Figure 2: Precision for each topic and service (relevance assessments per topic / total amount of single assessments), including standard error

In our experiment the Solr standard relevance ranking algorithm (based on TF-IDF) was taken as the baseline to which each of the three value-added services proposed had to compare. The average precision of TF-IDF over all topics and assessors was 57%, where values range between 17% and 75%. Ignoring the 17% value all other values are stable around the baseline. For two topics (83 and 84) the baseline was also the best judged result.

STR used the same SOLR ranking mechanism as the baseline but with the addition of automatically expanded queries. By expanding the query with the $n=4$ suggested terms with the highest confidence the average precision could be raised from 57 to 67% which is an impressive improvement in precision, despite that fact that a larger recall is obtained (due to OR-ing the query with the four additional terms). In three cases precision drops below the baseline (topic 84, 96 and 105). If we take standard error in consideration only topic 105 is a real outliner (45% vs. a baseline of 66%).

Looking at the four topics where STR was best (88, 110, 166 and173 with an improvement of 20% at least compared to the baseline) it can be seen that this positive query drift was because of the new perspective added by the suggested terms. STR added new key elements to the query that were not included before. For topic 88 ("Sports in Nazi Germany"), for instance, the suggested term with the highest confidence was "Olympic Games". A term that was not included in title or description in any way. Of course, sports in Nazi Germany is not only focused on the Olympic Games 1936, but with a high probability everything related to the Olympic Games 1936 had to do with sports in Nazi Germany and was in this way judged relevant by the assessors. Other topics showed comparable phenomena.

The two alternative re-ranking mechanisms Bradfordizing and Author Centrality achieved an average precision that is near the baseline (57%), namely 60% for Author Centrality and 57% for Bradfordizing. Author Centrality yielded a higher, but not significantly higher average precision than Solr as a conventional ranking mechanism[11]. Both re-rank mechanisms showed a stable behavior (again, expect some outlier, cp. Figure 2).

*Overlap of top document result sets*

However, a more important result as regards the two re-ranking services is that they point to quite other documents than other services. This indicates that the science models behind them provide a very different view to the document space.

---

[11] Moreover, we observed a high range of re-rankings done by Author Centrality. More than 90% of the documents in the result sets were captured by the author centrality based ranking.

A comparison of the intersection of the relevant top 10 document result sets between each pair of retrieval service shows that the result sets are nearly disjoint. 400 documents (4 services * 10 per service * 10 topics) only had 36 intersections in total (cp. Figure 3). Thus, there is no or very little overlap between the sets of relevant top documents obtained from different rankings.

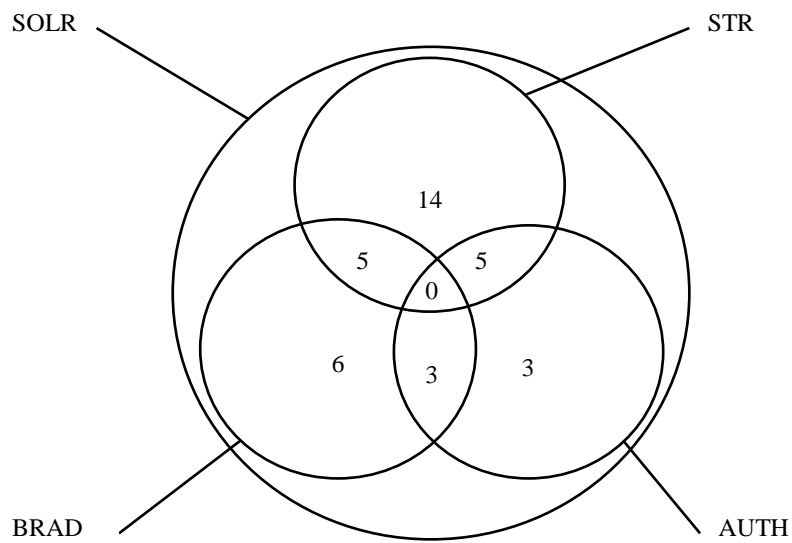

Figure 3: Intersection of suggested top n=10 documents over all topics and services (total of 400 documents)

AUTH and SOLR as well as AUTH and BRAD have just three relevant documents in common (for all 10 topics), and AUTH and STR have only five documents in common. BRAD and SOLR have six, and BRAD and STR have five relevant documents in common. The largest, but still low overlap is among SOLR and STR which have 14 common documents. That confirms that the models proposed provide views to the document space that differ greatly from standard retrieval models as well as from one another. This can be also seen as a positive validation of the adequacy of the science models taken for representing relevant and quite different scientific activities.

## Discussion

Two important implications emerge from the evaluation results: (1) The science models proposed provide beneficial effects to information retrieval quality. The precision tests turned out a precision of science-model driven search services which is at least as high as the precision of standard rankings. The more important effect of the models however is that they provide a particular view to the information space that is quite different from traditional retrieval methods such that the models open up new access paths to the knowledge space in question. (2) The science models studied are therefore verified as expressive models of science, as an evaluation of retrieval quality is seen as a litmus test of the adequacy of the models investigated. Moreover, it turned out that the results provided by the three science models investigated differ to a great extend which indicates that the models highlight very different dimensions of scientific activity. This also demonstrates that the models properly address the diversity of structures and dynamics in science.

The evaluation of retrieval quality achieved by a co-word model approach of query expansion, as performed by the STR, turned out significantly that a "query drift" (Mitra et al. 1998) towards terms that better reflect the scientific discourse(s) actually tends to retrieve more relevant documents (cp. Petras 2006). It is important to note that this positive effect is not only achieved by just mapping query terms to controlled terms from the indexing vocabulary, but mainly by linking the original query to the right research context the user has in mind, i.e. to research issues that are strongly co-related to the original term. The STR maps a query term to the cognitive structure of a field allowing the user to identify and select related topics and streams which obviously leads to more precise queries. Thus , the co-word model of science is verified as an expressive model of accessing the cognitive structure of a field and its various dimensions.

As regards the two re-ranking methods some added-values appear very clearly. On an abstract level, the re-ranking models can be used as a compensation mechanism for enlarged search spaces. Our analyses show that the hierarchy of the result set after re-ranking by Bradfordizing or Author Centrality is a completely different one compared to the original ranking. The user gets a new result cutout containing other relevant documents which are not listed in the first section of the original list. Additionally, the re-ranking via structure-oriented science models offer an opportunity to switch between term-based search and structure-oriented browsing of document sets (Bates 2002). On the other hand, modeling science into a core and a periphery – the general approach of both re-ranking models – always runs the risk and critic of disregarding important developments outside the core (cp. Hjorland and Nicolaisen 2005). Both models, however, imply the principle possibility to turn round the ranking in order to explicitly address publications of less central authors or publications in peripheral journals. Moreover, and probably more interesting, might be the ability of the models to point to items that appear between the top and the "midfield" of the structure, for instance publications in zone 2 journals or publications of "social climbers" in co-authorship networks (who seem to have a strong tendency of addressing innovative topics instead of mainstream issues, see Mutschke and Quan-Haase 2001).

Thus, it could be shown how structural models of science can be used to improve retrieval quality. The other way around, the IR experiment turned out that to the same extent to which science models contribute to IR (in a positive as well as negative sense), science-model driven IR might contribute to a better understanding of different conceptualizations of science (role of journals, authors and language in scientific discourses). Recall and precision values of retrieval results obtained by science model oriented search and ranking techniques seem to provide important indicators for the adequacy of science models in representing and predicting structural phenomena in science.

As regards the relationship between bibliometric-aided retrieval and traditional IR it turned out that, although the different perspectives aim at different objectives, on a generic level they share questions of the determination of the relevant information space and boundary setting in such a space. Thus, we could imagine that a future systematic comparison of bibliometric-aided retrieval and traditional IR approaches could be of relevance both for the questions "what is a scientific field?" as well as for "what is the scientific field relevant for my search?". In such a comparison, the different models of science could be explicitly addressed and compared, together with the different selection criteria as applied by the two retrieval approaches.

A further point that might be interesting from the perspective of science modeling is the degree of acceptance of science models as retrieval methods by the users of a scholarly IR system. The degree to which scientists are willing to use those models for finding what they are looking for (as particular search stratagems, as proposed by Bates 2002) are further relevant indicators for the degree to which the models intuitively meet the real research process. Thus, the major contributions of IR to science modeling might be to measure the expressiveness of existing science models and to generate novel models from the perspective of IR. In addition, the application and utilization of science model enhanced public retrieval systems can probably be a vehicle to better explain and communicate science and science models to a broader audience in the sense of public understanding of science.

However, a lot of research effort needs to be done to make more progress in coupling science modeling with IR. We see this paper as a first step in this area. The major challenge that we see here is to consider also the dynamic mechanisms which form the structures and activities in question and their relationships to dynamic features in scholarly information retrieval[12].

## Acknowledgement

We would like to express our grateful thanks to Andrea Scharnhorst for her valuable comments. Special thanks go to the students in two independent LIS courses at Humboldt University (guided by our former colleague Vivien Petras) and University of Applied Science in Darmstadt. These students took part in our first IRM retrieval test in the winter semester 2009/2010. We thank Hasan Bas who did the main implementation work for our assessment tool.
The project is funded by DFG, grant no. INST 658/6-1.

---

[12] See Huberman and Adamic 2004 and Mutschke 2004b for first attempts in that direction.